\documentclass[twocolumn]{aastex701}

\usepackage{amsmath,amssymb}
\usepackage{booktabs}
\usepackage{graphicx}
\usepackage{array}
\usepackage{xcolor}
\graphicspath{{./}}

\newcommand{\ellr}{\ell_R}
\newcommand{\ellx}{\ell_X}
\newcommand{\logm}{m}
\newcommand{\Msun}{M_\odot}
\newcommand{\pnd}{P_{\rm nd}}

\shorttitle{Forecasting IMBH Fundamental-Plane Tests}
\shortauthors{Santra et al.}

\begin{document}
\hypersetup{pdftitle={Forecasting Fundamental-Plane Tests of Intermediate-mass Black Holes in Radio-selected Dwarf Galaxies},pdfauthor={Ramananda Santra, Samik Mitra, Shahram Abbassi},pdfsubject={Magenta-highlighted revision responding to AAS79479 referee and data-editor reports}}

\title{Forecasting Fundamental-Plane Tests of Intermediate-mass Black Holes in Radio-selected Dwarf Galaxies}

\author[0009-0002-0373-570X]{Ramananda Santra}
\email{ramananda.santra@icts.res.in}
\affiliation{International Centre for Theoretical Sciences, Tata Institute of Fundamental Research, Bengaluru, Karnataka 560089, India}

\author[0000-0002-5730-0376]{Samik Mitra}
\email{samik.mitra@icts.res.in}
\correspondingauthor{Samik Mitra}
\affiliation{International Centre for Theoretical Sciences, Tata Institute of Fundamental Research, Bengaluru, Karnataka 560089, India}

\author[0000-0003-0428-2140]{Shahram Abbassi}
\email{sabbassi@uwo.ca}
\affiliation{Department of Physics and Astronomy, Western University, 1151 Richmond Street, London, Ontario N6A 3K7, Canada}

\begin{abstract}
A radio/X-ray Fundamental Plane (FP) test of an intermediate-mass black hole is informative only when the compact radio emission, hard X-rays, accretion state, and host association refer to the same accretor. We used published VLA, VLBA, \textit{Chandra}, and \textit{HST} observations of a radio-selected dwarf-galaxy sample to determine when present nondetections are constraining and what observations are required for a stronger test. The VLBA follow-up reveals that compactness alone does not establish association with the target dwarf, while for the best nuclear candidates the radio component relevant to the FP has not yet been isolated on the intermediate physical scales between the VLA and VLBA measurements. We therefore recast the FP as a forward question: for an assumed black-hole mass, how likely is a given radio observation to detect the compact core? The calculation includes intrinsic FP scatter, X-ray uncertainties, the radio spectral index, the accretion state, and source size relative to the beam. The existing VLBA nondetections have low completeness across the adopted intermediate-mass range and do not provide useful black-hole mass upper limits. Meaningful tests require deeper resolution-matched radio imaging, recovery of intermediate-scale structure, and contemporaneous measurements of the intrinsic hard-X-ray component. Because the calculation returns a mass-dependent detection function for each source, the same framework can be applied to larger dwarf-galaxy samples to quantify survey completeness and build selection functions for population studies of low-mass black holes.
\end{abstract}


\section{Introduction}
\label{sec:intro}

Massive black holes are ubiquitous in the nuclei of massive galaxies, but the formation of the first black-hole seeds and their subsequent growth remain open problems
\citep{KormendyHo2013, Inayoshi2020, Greene2020}. Current seed models span remnants of the first stars, massive objects formed by rapid gas collapse, and growth through stellar collisions and black-hole mergers in dense nuclear clusters. Recent work emphasises that these mechanisms can populate a continuous range of seed masses rather than a strict light--heavy dichotomy \citep{ReganVolonteri2024,Kritos2024}. Because later accretion and mergers can erase much of this initial information, dwarf galaxies provide an important laboratory for probing the low-mass black-hole population \citep{Mezcua2017,Greene2020}. Their black-hole occupation fraction and mass distribution can therefore constrain models of early seeding and growth
\citep{Miller2015,ReinesVolonteri2015,Greene2020}.

Identifying black holes in dwarf galaxies is nevertheless difficult. Direct dynamical measurements are generally restricted to the nearest systems, so most searches rely on optical emission lines, broad H$\alpha$, variability, X-rays, or radio emission
\citep{Reines2013,Baldassare2015,Baldassare2016,Baldassare2020}.  Each method has distinct contaminants and selection biases. Star formation can dominate optical observations, while X-ray binaries, ultraluminous X-ray sources (ULXs), and diffuse thermal gas can mimic or obscure weak nuclear X-ray emission \citep{Lemons2015,Miller2015,Nyland2017}. No single wavelength therefore provides an unambiguous census of low-mass black holes. High-resolution radio imaging offers a complementary diagnostic. Compact, high-brightness-temperature emission is difficult to explain through ordinary star formation and can reveal weak jets from accreting black holes
\citep{Condon1992,Ho2008,Reines2011,ReinesDeller2012}. Radio compactness alone does not establish association with the intended dwarf. Background Active Galactic Nuclei (AGNs) may be projected near the galaxy, while genuine massive black holes can remain displaced from the stellar center following mergers or inefficient dynamical sinking \citep{Bellovary2021,Reines2020}. Robust classification therefore requires both compact-source evidence and a secure host association.

These issues are central to the radio-selected dwarf-galaxy sample studied here and to the sequence of follow-up observations. For these candidates, however, the relevant radio component has not been isolated across the $\sim1$--100 pc interval between the VLBA and VLA measurements. Related work has used the FP with SKA sensitivity forecasts to assess the detectability of intermediate-mass black holes (IMBHs) in the globular-cluster system of NGC~1399 \citep{Karimi2024}; here we address the complementary problem of source-specific, resolution-aware completeness for nuclear candidates in dwarf galaxies. When a compact radio core and a compatible hard-X-ray component are both identified, the radio/X-ray Fundamental Plane (FP) provides a mass-sensitive description of sub-Eddington black-hole activity. A commonly used forward
form is
\begin{equation}
 \ellr=\xi_{RX}\ellx+\mu_{\rm FP}\logm+b_0,
 \label{eq:fp}
\end{equation}
where
$\ellr\equiv\log[\nu L_\nu(5\,{\rm GHz})/{\rm erg\,s^{-1}}]$,
$\ellx\equiv\log[L_{2-10\,{\rm keV}}/{\rm erg\,s^{-1}}]$, and
$\logm\equiv\log(M_\bullet/\Msun)$
\citep{Merloni2003,Falcke2004}. The relation is motivated by the approximate scale invariance of compact jets and radiatively inefficient accretion flows
\citep{Heinz2003,Kording2006,Yuan2014}. Its use for individual sources is nevertheless limited by substantial intrinsic scatter, accretion-state dependence, heterogeneous calibrations, non-simultaneous measurements, and the distinction between total and compact-core radio emission \citep{Gultekin2009,Plotkin2012,Gultekin2019}. In addition, a regression calibrated to predict radio luminosity cannot be inverted into an equivalent black-hole mass posterior without further assumptions. Recent work has further emphasized that an inverse-FP mass is physically meaningful only when the accretor identity, accretion state, and compact-core radio component are independently established \citep{Abbassi2026}.

We therefore do not interpret the current radio nondetections as inverse-FP mass measurements. Instead, we ask what sensitivity and angular resolution are required to detect a $10^5$--$10^6\,\Msun$ IMBH in the three strongest nuclear candidates. We use $10^6$--$10^7\,\Msun$ only as a transition and comparison interval.  The calculation treats map noise in linear flux density, propagates FP scatter and X-ray uncertainties, marginalises over radio spectral
index, and separates thermal-noise depth from angular-scale recovery. We then translate the resulting detection probabilities into observing requirements for present-day very long baseline interferometry (VLBI) and future high-sensitivity long-baseline facilities. The objective is not to infer a mass from the existing limits, but to define the observations required for an interpretable FP experiment. The paper is organized as follows. Section~\ref{sec:sample} describes the sample and its multiwavelength classification.  Section~\ref{sec:method} presents the forward FP detection model.  Section~\ref{sec:results} gives the current detection probabilities and observing requirements. Section~\ref{sec:discussion} discusses the source-specific follow-up strategy, and Section~\ref{sec:conclusions} summarizes the main conclusions and the broader applicability of the framework.

\section{Sample and Source Selection}
\label{sec:sample}

\citet{Reines2020} identified 13 compact 8--12 GHz sources whose radio luminosities exceed expectations from star formation. \citet{Sargent2022} observed all 13 with the VLBA at 8.67 GHz. Four were detected with high brightness temperatures, but they are also the four largest-offset radio sources. The other nine were not detected; such nondetections exclude only a sufficiently bright unresolved milliarcsecond core and do not rule out a fainter accreting black hole or emission on scales resolved out by the VLBA. \citet{Sturm2026} obtained \textit{Chandra} and \textit{HST} follow-up for 12 of the 13 sources; ID~48 was not included because an optical point source already suggested a background quasar. Five of the 12 have coincident X-ray and optical counterparts. ID~64 is a spectroscopically confirmed background AGN and ID~92 is consistent with a compact star-forming region plus an ultraluminous X-ray source, leaving IDs~26, 82, and 83 as the strongest nuclear accreting massive-black-hole candidates. This source-by-source sequence, summarized in Table~\ref{tab:branches}, is why the quantitative FP forecast focuses on those three objects. The 8.67 GHz VLBA observations of \citet{Sargent2022} reached rms levels of 21--59 $\mu$Jy beam$^{-1}$, and the four detected sources (IDs 2, 28, 48, and 65) have $T_B>10^6$ K. The largest recoverable angular scale was approximately 30 mas. Table~\ref{tab:branches} provides a one-row-per-source summary of the radio, X-ray, and optical follow-up and makes explicit why only three sources enter the quantitative forecast.

\begin{figure*}[t]
\centering
\includegraphics[width=0.92\textwidth]{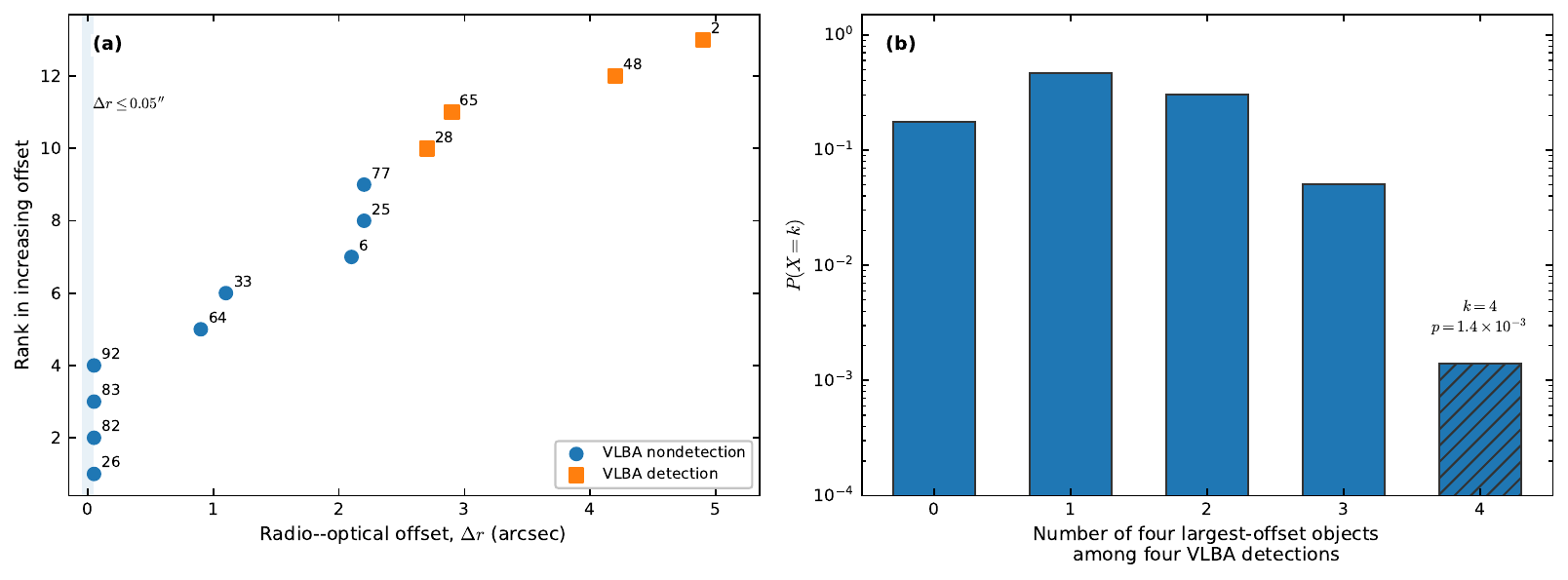}
\caption{\textit{Left:} Sources ordered by radio--optical offset; the shaded strip marks $\Delta r\leq0\farcs05$, and the
legend distinguishes VLBA detections from nondetections. \textit{Right:} Exact conditional distribution for the number of four largest-offset objects among four detections. The highlighted observed outcome is $k=4$, with $p=1.40\times10^{-3}$.}
\label{fig:offset}
\end{figure*}

The four detections are exactly the four largest-offset radio sources. Under the conditional null in which four detections are assigned randomly among 13 ranked offsets,
\begin{equation}
 p=\binom{13}{4}^{-1}=1.40\times10^{-3}.
 \label{eq:rank}
\end{equation}
This is a sample-level configuration probability, not a per-source background probability. We order the 13 sources by radio--optical offset and show that all four VLBA detections occupy the large-offset end of
the distribution (Fig.~\ref{fig:offset}), along with their corresponding exact conditional probability distribution. The observed outcome, in which all four detections are also the four largest-offset sources, has $p=1.40\times10^{-3}$.  This sample-level indication of contamination is not assigned to any individual source as a background probability.

The multiwavelength classifications in Table~\ref{tab:branches} follow \citet{Sargent2022}, \citet{Sturm2026}, and \citet{Eberhard2026}. The ``priority next observation'' is the measurement most likely either to establish whether the radio source is physically associated with the target dwarf or, for the nuclear candidates, to isolate the compact radio/hard-X-ray pair required for an interpretable FP test.

\begin{table*}[t]
\centering
\caption{Source-by-source summary of the 13 radio-selected candidates.}
\label{tab:branches}
\begingroup
\scriptsize
\setlength{\tabcolsep}{2.2pt}
\renewcommand{\arraystretch}{0.92}
\begin{tabular}{@{}ccllll@{}}
\toprule
ID & $\Delta r$ & \parbox[t]{0.10\textwidth}{VLBA} & \parbox[t]{0.10\textwidth}{\textit{Chandra}/\textit{HST}} & \parbox[t]{0.27\textwidth}{Present interpretation} & \parbox[t]{0.275\textwidth}{Priority next step} \\
\midrule
2  & 4.9\arcsec & \parbox[t]{0.10\textwidth}{Detected, high $T_B$} & \parbox[t]{0.10\textwidth}{No / no} & \parbox[t]{0.27\textwidth}{Compact nonthermal; association disfavored} & \parbox[t]{0.275\textwidth}{Counterpart redshift} \\
6  & 2.1\arcsec & \parbox[t]{0.10\textwidth}{Nondetection} & \parbox[t]{0.10\textwidth}{No / no} & \parbox[t]{0.27\textwidth}{Offset source; identity unresolved} & \parbox[t]{0.275\textwidth}{Deeper X-ray + optical/IR; 10--100 pc radio} \\
25 & 2.2\arcsec & \parbox[t]{0.10\textwidth}{Nondetection} & \parbox[t]{0.10\textwidth}{No / no} & \parbox[t]{0.27\textwidth}{Offset source; identity unresolved} & \parbox[t]{0.275\textwidth}{Deeper X-ray + optical/IR; 10--100 pc radio} \\
26 & $\leq0.05\arcsec$ & \parbox[t]{0.10\textwidth}{Nondetection} & \parbox[t]{0.10\textwidth}{Yes / yes} & \parbox[t]{0.27\textwidth}{Nuclear AGN; hard-X-ray decomposition uncertain} & \parbox[t]{0.275\textwidth}{Simultaneous VLBI + diagnostic hard X-rays} \\
28 & 2.7\arcsec & \parbox[t]{0.10\textwidth}{Detected, high $T_B$} & \parbox[t]{0.10\textwidth}{No / no} & \parbox[t]{0.27\textwidth}{Compact nonthermal; association disfavored} & \parbox[t]{0.275\textwidth}{Counterpart redshift} \\
33 & 1.1\arcsec & \parbox[t]{0.10\textwidth}{Nondetection} & \parbox[t]{0.10\textwidth}{No / no} & \parbox[t]{0.27\textwidth}{Offset source; identity unresolved} & \parbox[t]{0.275\textwidth}{Deeper X-ray + optical/IR; 10--100 pc radio} \\
48 & 4.2\arcsec & \parbox[t]{0.10\textwidth}{Detected, high $T_B$} & \parbox[t]{0.10\textwidth}{Not observed} & \parbox[t]{0.27\textwidth}{Optical point source suggests background quasar} & \parbox[t]{0.275\textwidth}{Counterpart redshift} \\
64 & 0.9\arcsec & \parbox[t]{0.10\textwidth}{Nondetection} & \parbox[t]{0.10\textwidth}{Yes / yes} & \parbox[t]{0.27\textwidth}{Background AGN at $z=0.761$} & \parbox[t]{0.275\textwidth}{Analyze at the background redshift} \\
65 & 2.9\arcsec & \parbox[t]{0.10\textwidth}{Detected, high $T_B$} & \parbox[t]{0.10\textwidth}{No / no} & \parbox[t]{0.27\textwidth}{Compact nonthermal; association disfavored} & \parbox[t]{0.275\textwidth}{Counterpart redshift} \\
77 & 2.2\arcsec & \parbox[t]{0.10\textwidth}{Nondetection} & \parbox[t]{0.10\textwidth}{No / no} & \parbox[t]{0.27\textwidth}{Offset source; identity unresolved} & \parbox[t]{0.275\textwidth}{Deeper X-ray + optical/IR; 10--100 pc radio} \\
82 & $\leq0.05\arcsec$ & \parbox[t]{0.10\textwidth}{Nondetection} & \parbox[t]{0.10\textwidth}{Yes / yes} & \parbox[t]{0.27\textwidth}{Nuclear AGN; hard X-ray + optical AGN signatures} & \parbox[t]{0.275\textwidth}{Simultaneous VLBI + hard X-rays} \\
83 & $\leq0.05\arcsec$ & \parbox[t]{0.10\textwidth}{Nondetection} & \parbox[t]{0.10\textwidth}{Yes / yes} & \parbox[t]{0.27\textwidth}{Nuclear massive-black-hole candidate} & \parbox[t]{0.275\textwidth}{Simultaneous VLBI + hard X-rays} \\
92 & $\leq0.05\arcsec$ & \parbox[t]{0.10\textwidth}{Nondetection} & \parbox[t]{0.10\textwidth}{Yes / yes} & \parbox[t]{0.27\textwidth}{Compact star-forming region + ULX} & \parbox[t]{0.275\textwidth}{No FP follow-up unless classification changes} \\
\bottomrule
\end{tabular}
\par\vspace{2pt}
\begin{minipage}{0.96\textwidth}
\scriptsize
\textit{Note.} Offsets are the radio--optical separations used in Figs.~\ref{fig:offset} and \ref{fig:diagnostics}. ``No / no'' indicates no significant X-ray or HST counterpart at the radio position in the follow-up analysis. ID~48 was not included in the \textit{Chandra}/\textit{HST} campaign of \citet{Sturm2026}. A VLBA nondetection constrains only a sufficiently bright unresolved core and is not by itself evidence against an accreting black hole.
\end{minipage}
\endgroup
\end{table*}

The three nuclear objects retained for forecasting have qualitatively stronger evidence than the rest of the sample. ID~26 has nuclear optical AGN diagnostics, including broad H$\alpha$, together with coincident X-ray and optical emission, although its hard-X-ray component is spectrally uncertain \citep{Reines2013,Baldassare2017,Sturm2026}. ID~82 has a hard nuclear X-ray source and independent AGN evidence from coronal [Fe~X], enhanced [O~I], and broad H$\alpha$ emission \citep{Molina2021,Sturm2026}. ID~83 has spatially coincident nuclear radio, X-ray, and optical emission \citep{Sturm2026}. The other objects either lack nuclear multiwavelength counterparts, are probable background sources, or are consistent with star formation and a ULX.

\begin{figure*}[t]
\centering
\includegraphics[width=0.96\textwidth]{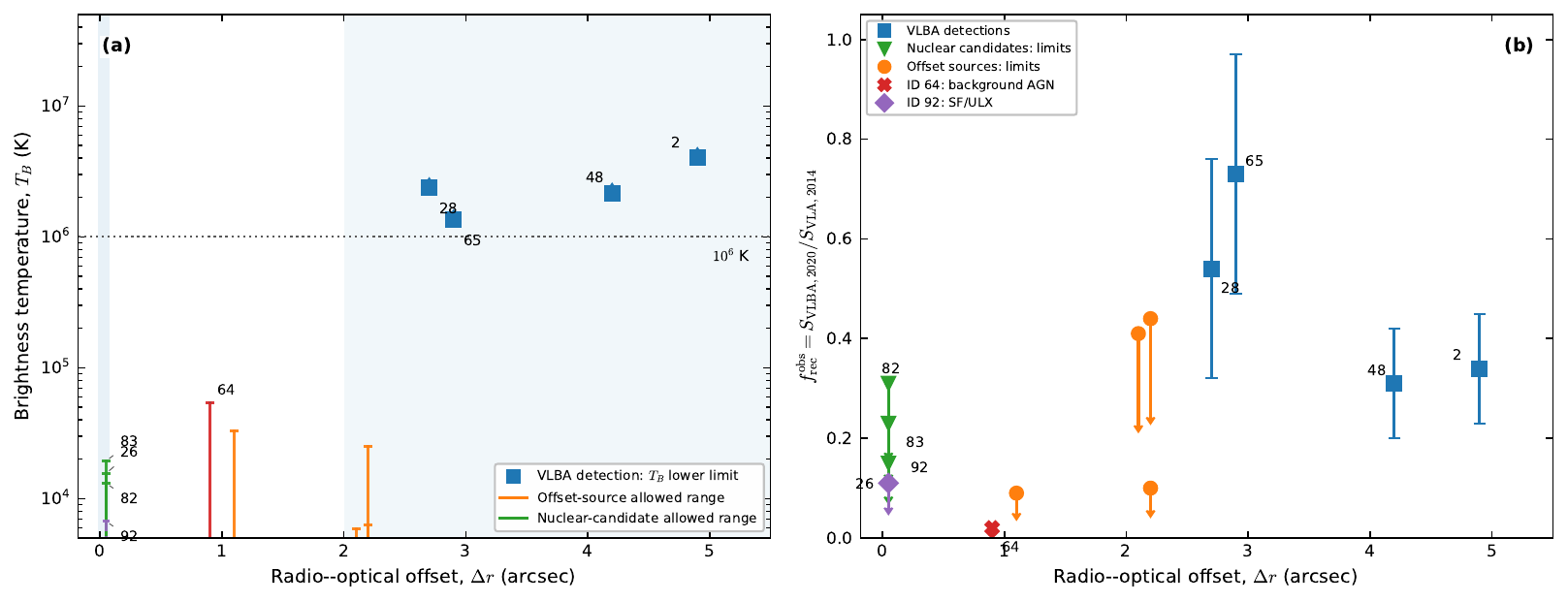}
\caption{\textit{Left:} Brightness-temperature constraints versus radio--optical offset for the 13-source sample, with the shaded regions marking the positionally nuclear and large-offset regimes is shown. A high $T_B$ establishes compact nonthermal emission but not association with the intended dwarf; vertical segments for nondetections are allowed ranges, not measurements; their bounds correspond to the VLA resolution limit and the VLBA largest recoverable scale. \textit{Right:} Non-simultaneous VLBA-to-VLA flux ratio is shown. Because the
epochs differ by about six years, $f_{\rm rec}^{\rm obs}$ mixes spatial filtering and variability. Data points use the measured catalogue offsets; only text labels are displaced.  Upper-limit arrow lengths are schematic.}
\label{fig:diagnostics}
\end{figure*}

The radio measurements provide two complementary constraints on source classification. The VLBA detections reach $T_B>10^6$ K, confirming compact nonthermal emission, but all occur at large radio--optical offsets (Fig.~\ref{fig:diagnostics}a).  Their compactness therefore establishes the nature of the radio source, not its association with the intended dwarf
galaxy. By contrast, the positionally nuclear candidates remain undetected on milliarcsecond scales, so the present data permit compact cores below the VLBA sensitivity as well as emission distributed over the unresolved intermediate scales between the VLA and VLBA observations. The VLBA-to-VLA flux ratios provide an independent view of this scale
dependence (Fig.~\ref{fig:diagnostics}b). The detected sources recover only a fraction of the earlier VLA flux, while the nondetections place upper limits on the milliarcsecond component. Because the two observations are separated
by approximately six years, the flux ratios cannot be interpreted as pure resolved-out fractions: they combine angular-scale filtering with possible intrinsic variability.  Together, the brightness-temperature and flux-recovery
constraints show that neither a VLBA detection nor nondetection is sufficient on its own; source association and intermediate-scale radio structure remain essential to identifying the component relevant for an FP analysis.

\section{Predictive Detection Model} \label{sec:method}

The model asks a forward observational question rather than assigning a mass to a radio upper limit. For an assumed black-hole mass, we first combine the measured hard-X-ray luminosity with an FP calibration to obtain a distribution of intrinsic compact-core radio luminosities. We then convert each radio luminosity to a 9 GHz flux density, allow for spectral-index uncertainty and beam dilution if the source is resolved, and pass the resulting peak flux through the map-noise detection threshold. Integrating over these quantities gives the probability that a source of the assumed mass would remain undetected. Finally, we solve the same calculation in reverse for the map rms required to reach a chosen detection completeness. This ordering keeps measurement noise, intrinsic FP scatter, and angular resolution as distinct ingredients.

The image noise is approximately Gaussian in linear flux density. For a
threshold $n_\sigma\sigma$, the probability of failing to detect a source of
mass $m$ is
\begin{align}
 \pnd(m)=&\int d\alpha\,d\ellx\,d\ellr\;
 p(\alpha)p(\ellx)
 \mathcal{N}(\ellr;\bar\ellr,\sigma_{R}) \nonumber\\
 &\times\Phi\!\left[
 \frac{n_\sigma\sigma-\eta_\theta S_9(\ellr,\alpha)}{\sigma}
 \right],
 \label{eq:pnd}
\end{align}
where $\bar\ellr=\xi_{RX}\ellx+\mu_{\rm FP}m+b_0$ and $\Phi$ is the
standard-normal cumulative distribution. In $\mu$Jy, the radio-luminosity to
flux-density mapping is
\begin{equation}
 S_9(\ellr,\alpha)=
 \frac{10^{\ellr}}{4\pi D_L^2(5\times10^9\,{\rm Hz})}
 \left(\frac{9}{5}\right)^\alpha 10^{29},
 \label{eq:fluxmap}
\end{equation}
where $D_L$ is in cm and the final factor converts cgs flux density to
$\mu$Jy; the small radio $K$-correction is neglected at the sample redshifts.
The source-size factor for a circular Gaussian is
\begin{equation}
 \eta_\theta=\frac{\theta_b^2}{\theta_b^2+\theta_s^2},
 \label{eq:size}
\end{equation}
with $\theta_b$ the geometric-mean restoring beam and $\theta_s$ the source
FWHM. Our baseline takes $p(\alpha)$ to be uniform on $[-0.7,+0.3]$; we also test a flat-core prior restricted to $[-0.2,+0.3]$. The published
asymmetric X-ray intervals are 90\% confidence ranges; for the predictive
integrals we approximate them by two-sided Gaussians with
$\sigma_\pm=\Delta_{90,\pm}/1.645$.  Equation~\ref{eq:pnd} is a
mass-prior-free, model-conditional predictive probability.

We next specify the astrophysical distributions entering Equation~\ref{eq:pnd}. Our baseline uses the \citet{Merloni2003} coefficients
$(\xi_{RX},\mu_{\rm FP},b_0)=(0.60,0.78,7.33)$ and
$\sigma_R=0.88$ dex. We marginalize over the converted X-ray likelihood approximations and a uniform $-0.7\leq\alpha\leq+0.3$, allowing steep, flat, and mildly inverted cores. For ID 26 we carry two distinct X-ray products: the photometric hard-band estimate, $\ellx=40.1_{-0.9}^{+0.4}$, and the spectral-model hard
component, $\ellx=39.0\pm0.4$; all quoted intervals here are 90\% confidence ranges and each scenario has its own Gaussian-equivalent uncertainty. As a regression-direction sensitivity check, we also rearrange the direct mass predictor of \citet{Gultekin2019},
\begin{equation}
 \mu=0.55+1.09R-0.59X,
 \label{eq:gultekin}
\end{equation}
where $\mu=\log(M_\bullet/10^8\Msun)$, $R=\log(L_R/10^{38})$, and
$X=\log(L_X/10^{40})$. Solving Equation~\ref{eq:gultekin} for $R$ gives $R=(\mu-0.55+0.59X)/1.09$, so its 0.96 dex scatter in $\mu$ maps to $\sigma_R=0.96/1.09=0.88$ dex in the radio direction. We use this rearranged G\"ultekin relation only as a check on sensitivity to regression direction. The resulting span
between the Merloni-based and G\"ultekin-based predictions is therefore not a confidence interval or a formal uncertainty on either FP calibration.

The rms required for a target detection probability $p$ satisfies
\begin{equation}
 1-\pnd(m_t;\sigma_p)=p.
 \label{eq:forecast}
\end{equation}
Here $m_t$ is the target black-hole mass and $\sigma_p$ is the map rms that yields detection probability p. We show $n_\sigma=3$, 4, and 5 because a known-position fit can use a different
false-positive threshold from a blind survey. The FP is most directly motivated for low/hard-state, radiatively inefficient accretion with a compact jet. We therefore adopt $\lambda_{\rm Edd}<0.03$ as an operational consistency criterion. This value lies within the $\sim1$--4\% Eddington range over which soft-to-hard transitions are observed in black-hole X-ray binaries \citep{MaccaroneState2003} and is consistent with the few-percent regime in which hot, radiatively inefficient flows are commonly expected \citep{Yuan2014}. It is not treated as a universal state boundary, which is why we explicitly test lower and higher choices. Writing $L_{bol}$ = $\kappa_X L_X$, we adopt 10 $\leq \kappa_X \leq 20$, representative of 2–10 keV bolometric corrections for low-Eddington-ratio AGNs \citep{Vasudevan2007}. With this range, the criterion corresponds to $log(L_X/L_Edd) \sim -2.8$ to $-2.5$, comparable to the scale identified from black-hole X-ray binaries by \citet{Abbassi2026}, who also show that a low Eddington ratio alone does not guarantee that the FP applies. The low-state criterion then requires:

\begin{equation}
 m>\ellx+\log\kappa_X-\log(0.03\times1.26\times10^{38}).
 \label{eq:state}
\end{equation}
Rather than applying only a central-value cut, we average the state condition
over the asymmetric $L_X$ distribution and a uniform $10\leq\kappa_X\leq20$,
\begin{align}
 P_{\rm state}(m)=&\int d\ellx\,d\kappa_X\,
 p(\ellx)p(\kappa_X) \nonumber\\
 &\times H\!\left[m-m_{\lambda}(\ellx,\kappa_X)\right],
 \label{eq:pstate}
\end{align}
where $H$ is the Heaviside function, and $m_{\lambda}(\ellx,\kappa_X)$ is the right-hand side of Eq.~\ref{eq:state}.  Forecasts are shown as solid where
$P_{\rm state}\geq0.95$, dashed through the transition, and dotted where
$P_{\rm state}\leq0.05$; for ID 26 we conservatively combine both X-ray
decompositions. Changing the adopted transition from $\lambda_{\rm Edd}=0.03$ to 0.01 or 0.05 shifts the corresponding mass boundary by $+0.48$ or $-0.22$ dex, respectively, without changing the radio detection probabilities.

The median point-source scaling follows directly from Equation~\ref{eq:fp},
\begin{equation}
 S_9\propto D_L^{-2}L_X^{\xi_{RX}}M_\bullet^{\mu_{\rm FP}},
 \label{eq:scaling}
\end{equation}
so the required rms has the same leading dependence. This scaling makes the framework portable to other dwarf-galaxy samples while the full calculation retains the scatter, measurement errors, spectral index, and detection threshold. The forecasts are conditional on the source being observed
near its currently inferred X-ray state. A shift
$\Delta\log L_X=\pm0.3$ ($\pm0.5$) dex changes the median expected radio flux by a factor $10^{0.60|\Delta\log L_X|}=1.51$ (2.00), and therefore changes a thermal exposure estimate by as much as a factor 2.3 (4). Coordinated epochs
are thus part of the experiment rather than an optional refinement.

\section{Results}
\label{sec:results}

\begin{figure*}[t]
\centering
\includegraphics[width=\textwidth]{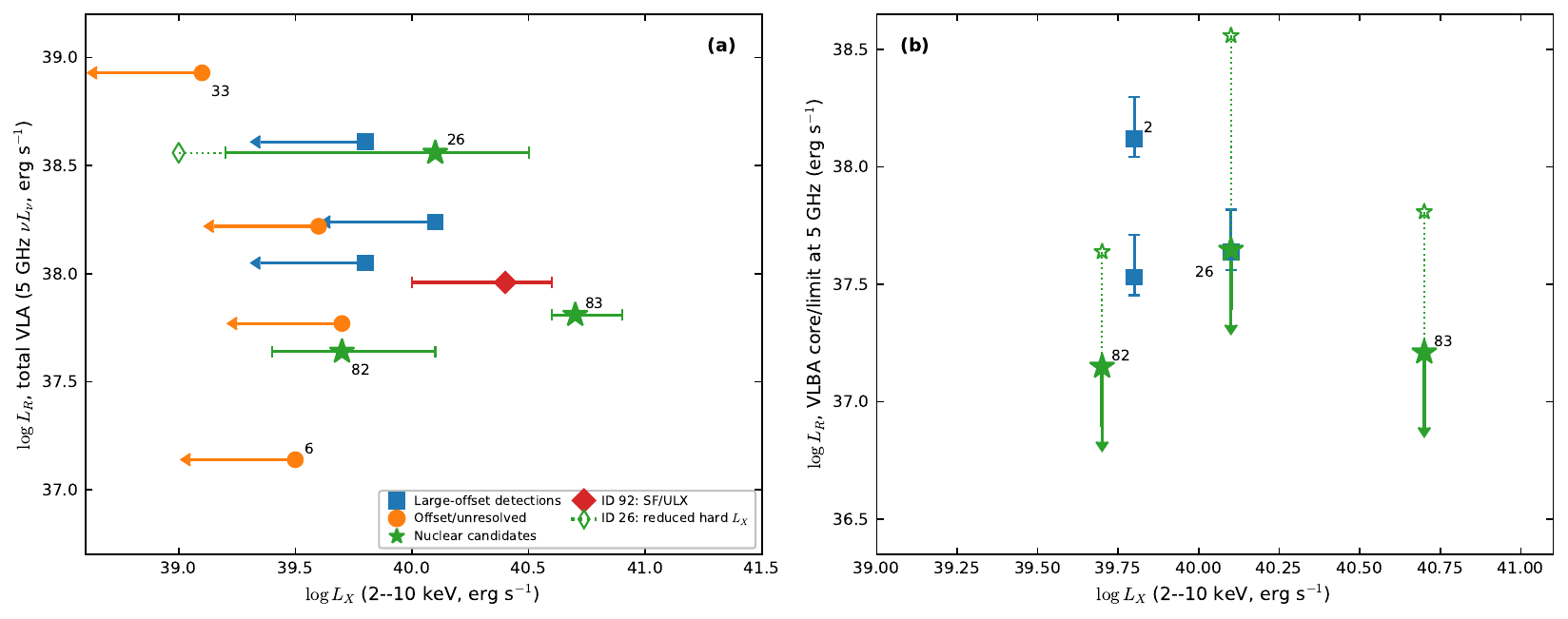}
\caption{Radio/X-ray coordinates before and after isolating the compact radio component. \textit{Left:} Total VLA radio luminosity at the target
dwarf-galaxy distance; leftward arrows denote X-ray upper limits, and the open diamond with dotted segment marks the alternative hard-$L_X$ estimate for
ID 26. \textit{Right:} VLBA detection or $3\sigma$ upper limit on the unresolved core, with vertical brackets spanning $-0.7\leq\alpha\leq+0.3$. Open stars mark the original VLA positions of IDs 26, 82, and 83, and dotted connectors indicate replacement of the VLA-scale by the VLBA-scale measurement; because the epochs are non-simultaneous, they are not measured temporal tracks. The VLBA limits do not exclude radio emission on
intermediate scales.}
\label{fig:components}
\end{figure*}

A source's position in the radio/X-ray plane depends strongly on which spatial component is assigned to the compact accretor. On the $\sim10^2$ pc scales sampled by the VLA, several sources have comparatively large radio-to-X-ray ratios (Fig.~\ref{fig:components}, left). Here ``radio bright'' means large $L_R$ at a given $L_X$, or equivalently large $L_R/L_X$; it does not mean that the absolute radio luminosity exceeds the X-ray luminosity. For IDs 26, 82, and 83, the total VLA luminosities give $log(L_R/L_X) \sim -1.5, -2.0$, and $-2.9$, well above the value $log R_X \sim -4.5$ commonly used to separate radio-loud from radio-quiet low-luminosity AGNs \citep{Terashima2003}, and 1.4–2.5 dex above the radio luminosity predicted for a $10^{6}$M$_{\odot}$ black hole at the measured $L_{X}$. Replacing the total VLA luminosity with the detected or censored milliarcsecond component shifts IDs 26, 82, and 83 downward by approximately 0.5--1 dex in radio luminosity (Fig.~\ref{fig:components}, right). The displacement denotes replacement of the VLA-scale measurement by the VLBA-scale measurement; because the epochs are non-simultaneous, it should not be interpreted as a measured temporal evolution. The VLA beams span 116--221 pc, the VLBA beams probe 0.6--1.8 pc, and the VLBA largest angular scale of 30 mas corresponds to 14--27 pc. The present observations therefore leave a broad interval of intermediate-scale radio structure
unconstrained.

Once the compact radio component is treated explicitly, the present VLBA nondetections provide little leverage on the adopted IMBH mass range. Across the regression-direction and X-ray sensitivity ranges, the current $5\sigma$ probabilities of detecting a $10^6\,\Msun$ point source are $4.9\times10^{-5}$--0.03, 0.03--0.06, and 0.08--0.17 for IDs 26, 82, and 83,
respectively. The existing limits therefore cannot be interpreted as astrophysically useful upper bounds on their black-hole masses.

\begin{figure*}[t]
\centering
\includegraphics[width=\textwidth]{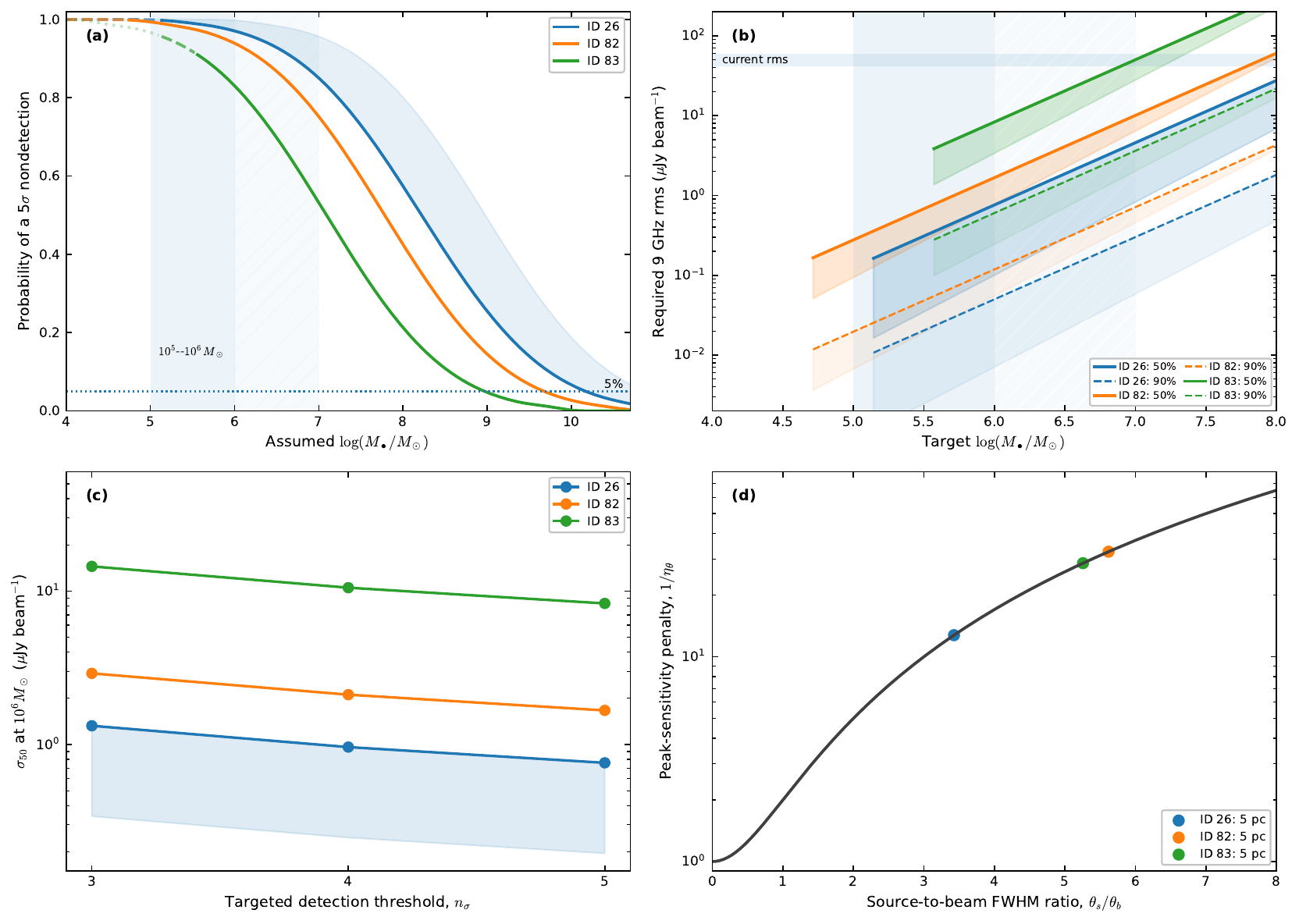}
\caption{\textit{Upper left:} Probability of a $5\sigma$ VLBA nondetection as a function of
assumed black-hole mass. Curve styles encode the marginalized low-Eddington-state validity: solid where $P_{\rm state}\geq0.95$, dashed through the transition region, and dotted where $P_{\rm state}\leq0.05$. The band for ID 26 spans the photometric and spectral-model hard-X-ray estimates. Solid shading marks the adopted $10^5$--$10^6\,\Msun$ IMBH interval, while hatching marks the $10^6$--$10^7\,\Msun$ comparison range. \textit{Upper right:} Point-source rms required for 50\% and 90\% completeness at a $5\sigma$ threshold. The shaded regions show the sensitivity to regression direction and, for ID 26, also span the two X-ray estimates; they are not confidence bands. \textit{Lower left:} Dependence of $\sigma_{50}$ on the
adopted detection threshold for a $10^6\,\Msun$ source. \textit{Lower right:} Facility-independent peak-sensitivity penalty as a function of the
source-to-beam FWHM ratio. Colored markers indicate the penalty for a 5 pc Gaussian source observed with the existing source-specific VLBA beams.}
\label{fig:forecast}
\end{figure*}

The sensitivity required for a meaningful test is summarized in
Fig.~\ref{fig:forecast}. At the current VLBA depths, the probability of a nondetection remains high throughout most of the $10^5$--$10^6\,\Msun$ interval. Reaching 50\% or 90\% completeness requires substantially deeper observations, with the exact requirement depending on the FP regression direction and, for ID 26, on the adopted hard-X-ray decomposition. The same calculation also separates the effects of detection
threshold and angular-size dilution from the nominal thermal-noise requirement.

Across the adopted sensitivity range, a 50\% chance of detecting a $10^6\,\Msun$ point source requires
$\sigma_{50}=0.10$--0.76, 0.77--1.67, and 3.37--8.31 $\mu$Jy beam$^{-1}$ for IDs 26, 82, and 83. The corresponding
90\% completeness ranges are 0.007--0.050, 0.055--0.118, and
0.244--0.602 $\mu$Jy beam$^{-1}$. The extreme 90\% values are diagnostics of  the low-radio tail produced by the nearly one-dex FP scatter, not realistic facility requirements or routine observing recommendations. A narrower flat-core prior, $-0.2\leq\alpha\leq+0.3$, raises both bounds of $\sigma_{50}$ by approximately 16\% for all three sources; this is smaller than the regression-direction and ID 26 X-ray-decomposition ranges. The adopted detection criterion changes the requirements at a similarly modest level. A known-position $3\sigma$ fit increases the allowable 50\% rms by a
factor of approximately 1.7 relative to a blind $5\sigma$ criterion. Source extent can impose a considerably larger penalty. For a circular Gaussian source,
\[
\frac{1}{\eta_\theta}
 = 1+\left(\frac{\theta_s}{\theta_b}\right)^2,
\]
so a 5 pc source (a representative size between the VLBA beam and its largest recoverable scale) observed with the existing VLBA restoring beams would reduce the peak flux by factors of approximately 13, 33, and 29 for IDs 26, 82, and 83. A future 1 mas image may suffer an even larger surface-brightness penalty,
whereas tapering to 5--10 mas can improve sensitivity to partially resolved emission. Deeper integration alone cannot recover structure that is resolved out by the chosen baselines.

\begin{deluxetable*}{lccccccc}
\tabletypesize{\footnotesize}
\tablewidth{0pt}
\tablecaption{Inputs and $10^6\,\Msun$ point-source forecasts.\label{tab:forecast}}
\tablehead{
\colhead{ID} & \colhead{$D_L$} & \colhead{$\log L_X$ (90\%)} &
\colhead{Beam; LAS} & \colhead{$\sigma_{\rm now}$} &
\colhead{$P_{\rm det,now}$} & \colhead{$\sigma_{50}$} & \colhead{$\sigma_{90}$}\\
\colhead{} & \colhead{(Mpc)} & \colhead{(erg s$^{-1}$)} &
\colhead{(pc)} & \colhead{($\mu$Jy beam$^{-1}$)} & \colhead{} &
\colhead{($\mu$Jy beam$^{-1}$)} & \colhead{($\mu$Jy beam$^{-1}$)}
}
\startdata
26 & 200 & $40.1_{-0.9}^{+0.4};\ 39.0\pm0.4$ & $1.76\times1.21;\,27$ & 41 & 4.9e-05--0.029 & 0.10--0.76 & 0.007--0.050 \\
82 & 113 & $39.7_{-0.3}^{+0.4}$ & $1.20\times0.66;\,16$ & 41 & 0.03--0.06 & 0.77--1.67 & 0.055--0.118 \\
83 & 101 & $40.7_{-0.1}^{+0.2}$ & $1.46\times0.62;\,14$ & 59 & 0.08--0.17 & 3.37--8.31 & 0.244--0.602 \\
\enddata
\tablecomments{The sensitivity ranges span the Merloni generative forecast and
the rearranged G\"ultekin regression-direction check; they are not confidence
intervals.  ID 26 also spans the photometric
$\ellx=40.1_{-0.9}^{+0.4}$ and spectral-model $\ellx=39.0\pm0.4$ products
(90\% intervals).  The predictive integrals divide the tabulated 90\%
intervals by 1.645 to obtain Gaussian-equivalent $1\sigma$ widths.  Beam and
largest angular scale (LAS) are both in pc; the LAS is derived from 30 mas.
The marginalized low-Eddington transitions are shown directly in
Fig.~\ref{fig:forecast}(a).}
\end{deluxetable*}

The facility estimates in Table~\ref{tab:facilities} translate these sensitivity requirements into approximate observing times. For the current VLBA X-band sensitivity of approximately
8 $\mu$Jy beam$^{-1}$ in 8 hr at 4 Gbps \citep{napier1994very, VLBAOSS2026}, simple $t^{-1/2}$ scaling implies that ID 83 can be tested
at approximately 50\% completeness with a deep VLBA or High Sensitivity Array program. This estimate remains conditional on the source being unresolved and should not be interpreted as a guaranteed detection forecast.

Matching the resolution of the radio observation is particularly important for the ngVLA comparison. At 8 GHz, the ngVLA reference design combines sub-$\mu$Jy sensitivity with continental-scale baselines \citep{Selina2018}. Current resolution-dependent performance estimates give $0.14\,\mu{\rm Jy\,beam^{-1}}$ in 1 hr with all-baseline natural weighting, $0.15\,\mu{\rm Jy\,beam^{-1}}$ at 10 mas, and $3.20\,\mu{\rm Jy\,beam^{-1}}$ at 1 mas
\citep{ngVLAPerformance}. Using the 1 mas value appropriate for a parsec-core test gives approximate thermal-noise estimates of 20--1000 hr for ID 26, 4--20 hr for ID 82, and 0.1--1 hr for ID 83. Practical observations will require additional time for phase referencing, calibration, flagging, beam control, and any resolved source structure. ID 82 is therefore a plausible but
nontrivial future long-baseline target, whereas ID 26 remains impractical unless its intrinsic hard $L_X$ is securely isolated and lies near the favorable end of the FP sensitivity range. Intermediate-baseline imaging remains necessary even when the required point source sensitivity is achieved. SKA-Mid alone will have baselines extending to approximately 150 km
\citep{Braun2019,Pellegrini2021}. At 8 GHz, the corresponding nominal
angular scale is approximately 50 mas, or 24--46 pc for the sources
considered here. SKA-Mid can therefore recover part of the missing VLA--VLBA scale range, whereas a parsec-scale FP measurement would require SKA-Mid
to participate as a phased element in a global VLBI network
\citep{Paragi2015,Garcia-Miro2019}.

The radio forecasts are also conditional on contemporaneous measurements of the intrinsic hard-X-ray component.  We target a precision of $\lesssim0.3$ dex in $L_X$: ID 26 requires separation of the thermal and hard components, ID 82 requires a contemporaneous luminosity and spectral slope, and ID 83 requires confirmation that the nuclear hard source persists.
Source-specific response simulations, rather than a universal count threshold, should determine the corresponding X-ray exposure requirements.

\begin{deluxetable*}{lcccc}
\tabletypesize{\scriptsize}
\tablewidth{0pt}
\tablecaption{Facility interpretation for 50\% completeness at $10^6\,\Msun$.\label{tab:facilities}}
\tablehead{
\colhead{ID} & \colhead{$\sigma_{50}$ range} & \colhead{VLBA-equivalent time} &
\colhead{ngVLA 1 mas time} & \colhead{Coordinated X-ray requirement}\\
\colhead{} & \colhead{($\mu$Jy beam$^{-1}$)} & \colhead{(hr)} & \colhead{(hr)} & \colhead{}
}
\startdata
26 & 0.10--0.76 & $\sim10^3$ to $>10^4$ (infeasible) & $\sim20$--1000 & \parbox[t]{0.35\textwidth}{separate thermal/hard components; $\sigma_{\log L_X}\lesssim0.3$ dex} \\
82 & 0.77--1.67 & $\sim200$--900 & 4--20 & \parbox[t]{0.35\textwidth}{contemporaneous hard $L_X$ and slope to $\lesssim0.3$ dex} \\
83 & 3.37--8.31 & 7--50 & 0.1--0.9 & \parbox[t]{0.35\textwidth}{simultaneous nuclear hard-X detection to $\lesssim0.3$ dex} \\
\enddata
\tablecomments{VLBA times scale from 8 $\mu$Jy beam$^{-1}$ in 8 hr and are reported only to order of magnitude.  ngVLA entries scale from the resolution-matched 1 mas sensitivity of 3.20 $\mu$Jy beam$^{-1}$ in 1 hr, not the all-baseline natural-weight value. They are thermal-noise lower bounds rather than proposal-ready exposures; phase referencing, flagging, beam requirements, calibration overheads, source structure, and minimum scheduling blocks will increase the practical time.  The final column specifies the required contemporaneous hard-X-ray product for an FP test to be interpretable.}
\end{deluxetable*}

\section{Discussion and Observing Strategy}
\label{sec:discussion}

The three priority sources require different experiments.  ID 26 has broad and narrow-line evidence for an accreting massive black hole \citep{Reines2013,Baldassare2017}, but its inferred hard X-ray luminosity is sensitive to the decomposition of thermal and nonthermal emission \citep{Sturm2026}. Its next observation should therefore first secure the intrinsic hard component, because the corresponding radio-sensitivity requirement spans almost an order of magnitude. ID 82 has high-ionization optical evidence and a hard nuclear \textit{Chandra} source \citep{Molina2021,Sturm2026}. Its resolution-matched ngVLA estimate of approximately 4--20 hr makes it a plausible but nontrivial coordinated long-baseline radio/X-ray target. ID 83 has the most favourable radio forecast. Under the point-source assumption and the adopted regression-direction sensitivity range, a sufficiently deep present-generation VLBI observation can reach approximately 50\% completeness for a $10^6\,\Msun$ source, while coordinated
optical and X-ray observations would strengthen its astrophysical classification.

A decisive experiment should use at least two radio resolutions. A milliarcsecond image tests for the compact high-$T_B$ core required for an FP measurement, whereas an intermediate-baseline image determines whether the VLA emission arises from a compact jet, a resolved outflow, or star formation on 10--100 pc scales. Radio and hard X-ray epochs should also be coordinated. The quoted radio depths are conditional on the currently inferred $L_X$; a 0.3--0.5 dex change in X-ray state can alter the required exposure by factors of approximately 2--4. Repeated observations should therefore measure radio/X-ray covariance rather than absorb all variability into the intrinsic FP scatter. Observational studies should report $P_{\rm det}(m)$ together with the angular scale, source-size assumption, host-association evidence, X-ray decomposition, and adopted detection threshold. A single inverse-FP mass estimate hides each of these physical and statistical choices.

The forecasts remain conditional on several assumptions. First, the FP must describe the relevant accretion state, and the compact radio and hard X-ray components must arise from the same accretor. Second, the adopted intrinsic scatter represents population-level variation and cannot distinguish among temporal variability, source-to-source differences, and departures from the assumed accretion-flow model. Third, the point-source forecasts are optimistic whenever the radio emission extends beyond the restoring beam. Fourth, the split-Gaussian representations of the published asymmetric X-ray intervals are approximations; posterior samples from the original low-count spectral analyses would be preferable when available. Finally, the adopted $\lambda_{\rm Edd}=0.03$ threshold is operational, and Figure~\ref{fig:forecast} encodes state validity separately from the radio detection calculation. These conditions do not invalidate the calculation; rather, they define when a radio nondetection becomes informative. A deep image with inappropriate angular-scale coverage, or without a contemporaneous hard-X-ray measurement, can remain insensitive to an IMBH even when its nominal thermal noise appears
sufficient.

The same forward formulation can be extended to substantially larger samples without assigning an inverse-FP mass to every source. Given a distance, hard-X-ray likelihood, radio beam and sensitivity, source-size model, and FP calibration, the calculation returns a mass-dependent detection function $P_{{\rm det},i}(m)$ for each galaxy. For a sample of $N$ objects, a useful survey-level summary is
\begin{equation}
 C(m)=\frac{1}{N}\sum_{i=1}^{N} P_{{\rm det},i}(m),
 \label{eq:samplecomp}
\end{equation}
where $C(m)$ is the mean completeness for an active, FP-compatible black hole of mass $m$. For example, $C=0.3$ in a 100-galaxy program means that about 30 radio detections would be expected if every target hosted an active black hole in the modeled state. The observed detection count can then be interpreted only after introducing an active fraction, host-association probabilities, and contaminating populations. This separates the instrumental selection function from the astrophysical occupation and duty-cycle problem. The same source-level probabilities can also rank follow-up targets and identify nondetections that were intrinsically uninformative. In a future radio/X-ray dwarf-galaxy survey, the set of $P_{{\rm det},i}(m)$ therefore provides the observational-completeness term needed for hierarchical constraints on the active low-mass black-hole population; converting those constraints into a total occupation fraction requires an additional accretion-duty-cycle model.

\section{Conclusions}
\label{sec:conclusions}

We recast the radio/X-ray Fundamental Plane from an inverse mass estimator as  a source-specific observing forecast for radio-selected massive black-hole candidates in dwarf galaxies. This distinction is essential: a radio
nondetection constrains an IMBH only when the relevant compact component, angular scale, hard-X-ray luminosity, accretion state, and host association are all specified consistently. Applying this framework to the published VLA, VLBA, \textit{Chandra}, and \textit{HST} data gives the following results summarised below:

\begin{enumerate}

\item The sample-level radio configuration cautions against interpreting compactness alone as evidence for a dwarf-galaxy black hole. The four high-$T_B$ VLBA detections are exactly the four largest-offset sources among the 13 objects, an outcome with conditional probability $p=1.40\times10^{-3}$. This does not assign a background probability to an individual source, but it shows that source association must be established before the compact radio luminosity can be used in an FP analysis. 

\item The existing VLBA nondetections have little statistical power in the adopted $10^5$--$10^6\,\Msun$ IMBH interval.  Across the X-ray and regression-direction sensitivity ranges, the current $5\sigma$ probabilities of detecting a $10^6\,\Msun$ point source are only $4.9\times10^{-5}$--0.03, 0.03--0.06, and 0.08--0.17 for IDs 26, 82, and 83. The absence of a VLBA detection therefore does not provide an astrophysically useful upper limit within the adopted $10^5$--$10^6\,\Msun$ mass interval.

\item A 50\% chance of detecting a $10^6\,\Msun$ point source requires 9 GHz rms sensitivities of 0.10--0.76, 0.77--1.67, and
3.37--8.31 $\mu$Jy beam$^{-1}$ for IDs 26, 82, and 83, respectively. The corresponding 90\% completeness requirements are substantially deeper because the nearly one-dex FP scatter produces a long low-radio-luminosity tail. These extreme depths should be interpreted as a limitation of an individual-object
FP experiment, not as routine observing recommendations.

\item The VLA measurements sample approximately 100 pc scales, whereas the existing VLBA beams probe approximately 1 pc and filter emission on intermediate scales. If the FP-related component has a 5 pc Gaussian FWHM, its peak flux in the present VLBA beams is reduced by factors of approximately 13, 33, and 29 for IDs 26, 82, and 83.  Increasing integration time cannot recover emission that has been resolved out; a successful experiment must combine milliarcsecond imaging with a tapered or intermediate-baseline measurement.

\item The three priority objects require different observing strategies. Under the point-source assumption and conditional on a contemporaneous hard X-ray state, ID 83 is the only source that can be tested at approximately 50\% completeness with a sufficiently deep present-generation VLBI or High Sensitivity Array observation. ID 26 is not presently radio limited in a simple
sense: its required depth spans nearly an order of magnitude because the intrinsic hard-X-ray component has not been securely isolated. Improved X-ray spectral decomposition must therefore precede an expensive radio experiment on this source.

\end{enumerate}

Beyond the three objects considered here, the same framework can be extended to substantially larger samples of radio- and X-ray-selected dwarf galaxies. For each source, the calculation requires only a distance, a hard-X-ray likelihood, a radio sensitivity and beam, an assumed source-size distribution, and an explicit FP calibration. It can therefore be incorporated into
survey-scale analyses to predict source-by-source completeness before observations are undertaken, to identify the candidates for which additional depth or angular resolution is most valuable, and to distinguish genuinely informative nondetections from observations that had little probability of detecting an IMBH in the first place. Applied uniformly across future
multiwavelength samples, these detection functions could also provide the selection model required for hierarchical studies of the low-mass black-hole population, including constraints on the active fraction and, with additional assumptions about accretion and source classification, the black-hole occupation fraction in dwarf galaxies.

\begin{acknowledgments}
We thank the anonymous referee for their constructive comments that have improved the overall clarity and presentation of the paper. R.S. and S.M. acknowledge support from the Department of Atomic Energy, Government of India, under project no. RTI4019. The authors acknowledge the use of generative-AI tools for language editing; all scientific content, calculations, references, and final text were checked by the authors.
\end{acknowledgments}

\software{Python, NumPy, SciPy, Matplotlib}

\begin{contribution}
R.S. and S.M. contributed equally to the scientific idea formulation, methodology, analysis, interpretation, and preparation of this manuscript.
S.A. contributed to the scientific interpretation and revision of
the manuscript. All authors reviewed and approved the final manuscript.
\end{contribution}

\section*{Data and Code Availability}
This work uses published radio, X-ray, and optical measurements and source classifications. We perform no new reduction or reanalysis of the VLA, VLBA, \textit{Chandra}, or HST observations. The derived products generated for this work, including the machine-readable input catalogue, source-level metadata, Fundamental-Plane forecast products, sensitivity calculations, and the scripts used to reproduce the figures and
tables, are publicly available at \citet{santra_2026_23100393} \dataset[doi:10.5281/zenodo.23100393]{https://zenodo.org/records/23100393}.

\bibliography{references}
\bibliographystyle{aasjournalv7}

\end{document}